\documentclass[%
 aip,
 amsmath,amssymb,
 reprint,%
]{revtex4-1}

\usepackage{graphicx}
\usepackage{dcolumn}
\usepackage{bm}

\usepackage[utf8]{inputenc}
\usepackage[T1]{fontenc}
\usepackage{mathptmx}
\usepackage{etoolbox}

\makeatletter
\def\@email#1#2{%
 \endgroup
 \patchcmd{\titleblock@produce}
  {\frontmatter@RRAPformat}
  {\frontmatter@RRAPformat{\produce@RRAP{*#1\href{mailto:#2}{#2}}}\frontmatter@RRAPformat}
  {}{}
}%
\makeatother
\begin{document}

\preprint{AIP/123-QED}

\title{Iterative charge equilibration for fourth-generation high-dimensional neural network potentials}

\author{Emir Kocer}
\affiliation{Lehrstuhl f\"ur Theoretische Chemie II, Ruhr-Universit\"at Bochum, 44780 Bochum, Germany}
\affiliation{Research Center Chemical Sciences and Sustainability, Research Alliance Ruhr, 44780 Bochum, Germany}
\author{Andreas Singraber}
\affiliation{University of Vienna, Faculty of Physics, Boltzmanngasse 5, A-1090 Vienna, Austria}
\author{Jonas A. Finkler}
\affiliation{Department of Chemistry and Bioscience, Aalborg University, 9220 Aalborg,
Denmark}
\author{Philipp Misof}
\affiliation{University of Vienna, Faculty of Physics, Boltzmanngasse 5, A-1090 Vienna, Austria}
\author{Tsz Wai Ko}
\affiliation{Aiiso Yufeng Li Family Department of Chemical and Nano Engineering, UC San Diego, 9500 Gilman Dr, La Jolla, CA 92093-0448, United States}
\author{Christoph Dellago}
\affiliation{University of Vienna, Faculty of Physics, Boltzmanngasse 5, A-1090 Vienna, Austria}
\author{J\"{o}rg Behler}
\email{joerg.behler@rub.de}
\affiliation{Lehrstuhl f\"ur Theoretische Chemie II, Ruhr-Universit\"at Bochum, 44780 Bochum, Germany}
\affiliation{Research Center Chemical Sciences and Sustainability, Research Alliance Ruhr, 44780 Bochum, Germany}

\date{\today}

\begin{abstract}
Machine learning potentials (MLP) allow to perform large-scale molecular dynamics simulations with about the same accuracy as electronic structure calculations provided that the selected model is able to capture the relevant physics of the system. For systems exhibiting long-range charge transfer, fourth-generation MLPs need to be used, which take global information about the system and electrostatic interactions into account. This can be achieved in a charge equilibration (QEq) step, but the direct solution (dQEq) of the set of linear equations results in an unfavorable cubic scaling with system size making this step computationally demanding for large systems. In this work, we propose an alternative approach that is based on the iterative solution of the charge equilibration problem (iQEq) to determine the atomic partial charges. We have implemented the iQEq method, which scales quadratically with system size, in the parallel molecular dynamics software LAMMPS for the example of a fourth-generation high-dimensional neural network potential (4G-HDNNP) intended to be used in combination with the n2p2 library. The method itself is general and applicable to many different types of fourth-generation MLPs. An assessment of the accuracy and the efficiency is presented for a benchmark system of FeCl$_3$ in water.
\end{abstract}

\maketitle

\section{Introduction}\label{sec:introduction}

The accurate description of potential energy surfaces (PES) in an efficient way remains one of the central challenges in computational chemistry and materials science. Established electronic structure methods, such as density functional theory (DFT)~\cite{P3910}, offer a high accuracy but are prohibitively expensive for large systems. To address this challenge, machine learning potentials (MLPs)~\cite{P4885, P6121, P5673,P5793,P6102,P6112,P6131} have emerged as a powerful alternative, offering an accuracy comparable to first-principles methods at strongly reduced computational costs, making them a useful tool for atomistic simulations. Since the first MLP has been proposed in 1995~\cite{P0316}, significant advances have been made in the development of MLPs, which may now, for instance, be classified into four generations~\cite{P5977}.

Early first-generation MLPs are applicable to low-dimensional systems and typically employ feed-forward neural networks to describe to potential energy of the system \cite{P0316,P0421,P1388,P0830}. The extension of MLPs to large condensed systems containing thousands of atoms became possible with the second generation of MLPs, which is based on representing the energy of the system as a sum of environment-dependent atomic energies. The first example of this generation has been high-dimensional neural network potentials (HDNNP) introduced by Behler and Parrinello in 2007 \cite{P1174}. To distinguish these HDNNPs from extended versions introduced in the following years, nowadays often the term 2G-HDNNP is used for these HDNNPs. Today, many very accurate MLPs of this generation are available~\cite{P1174,P2630,P4644,P4862,P5794,P4945,P5886}. Moreover, the introduction of MLPs employing modern message passing neural networks~\cite{P5368} enables the extension of the atomic environments by passing information from atom to atom~\cite{P5366,P5817,P6017,P6572}. Still, these potentials remain (semi-)local due to the limited number of passing steps employed in practical applications.

In spite of many successful applications, second-generation MLPs rely on environment-dependent atomic energies only and therefore do not incorporate any long-range interactions. This gap has been filled by the third generation that extends MLPs by modeling the electrostatic energy and forces through environment-dependent atomic charges \cite{P2391,P2962,P3132,P5577,P5313,P5629,P6272,P6200}. While these potentials offer some improvements in accuracy for some systems~\cite{P5975}, the computational cost is often increased by the need for computing machine learned charges and the explicit calculation of electrostatics, e.g. via an Ewald sum~\cite{P0238,P0935}, which can become the bottleneck in large-scale molecular dynamics (MD) simulations. As a comment, it should be noted that the representation of charges and electrostatic multipoles by machine learning has been suggested as early as 2007~\cite{P2391} to improve the accuracy of the electrostatic energy in classical force fields, i.e., at about the same time as the introduction of second generation MLPs~\cite{P1174} according to the classification scheme employed here.

Despite the inclusion of long-range electrostatics in third-generation MLPs, these potentials are still ``local'' in the sense that they exclusively make use of environment-dependent properties such as atomic energies and charges. As a consequence, they are unable to describe dependencies on distant changes in a system such as (de)protonation reactions~\cite{P5932} or electron transfer in redox processes~\cite{P6302}, which may result in long-range charge transfer within the system. Such \textit{nonlocal} phenomena are included in fourth-generation MLPs, in which the atomic charges depend on the entire system. 

The Charge Equilibration Neural Network Technique (CENT) \cite{P4419} was the first MLP to incorporate long-range charge transfer by using atomic electronegativities as environment-dependent properties learned via neural networks. These electronegativities are then used in a global charge equilibration (QEq) \cite{P1448} step to predict the charges of all atoms in the system~\cite{P4990,P5864,P5203}. In recent years, fourth-generation MLPs have become an active field of research and several methods have been proposed~\cite{P5932,P6666,P6122,P6310,P6542,P5859,P6829}.
An important example are fourth-generation high-dimensional neural network potentials (4G-HDNNP)~\cite{P5932}, which combine the advantages of CENT, i.e., global charges and electrostatics determined from QEq, and of 2G-HDNNPs, i.e., the accurate description of arbitrary local interactions within the atomic environments. Like in CENT, the QEq step in 4G-HDNNPs makes use of environment-dependent electronegativities expressed by atomic neural networks, but unlike CENT, the training of these neural networks does not aim to directly minimize the electrostatic energy. Instead, the electronegativities are optimized to match a set of reference charges  derived from DFT calculations. As these charges are determined from all electronegativities, they account for the structure and composition of the entire system, which ensures a globally correct distribution of charges. It is important to note that the equilibration of the charge density affects also local bonding. Therefore, the atomic charges serve -- next to determining long-range electrostatics -- also as additional global descriptors to compute the atomic energies in combination with atom-centered symmetry functions~\cite{P2882}, which provide a local structural fingerprint of the environments.

A disadvantage of the \textit{direct} charge equilibration step (dQEq) used in CENT and 4G-HDNNPs is the cubic scaling of solving the underlying matrix equations with respect to the number of atoms, which becomes computationally demanding for large systems. These costs thus restrict the system size in molecular dynamics (MD) simulations, which require updated atomic charges at each new step. In a very recent study, Gubler et al.~\cite{P6789} reformulated the QEq problem to avoid the explicit computation of coefficient matrix elements leading to a quasi-linear scaling of the method. In this work, we adopt a similar strategy and present a modified 4G-HDNNP method that builds upon the original method, but focuses on an efficient solution for the charge and force calculation. This improved efficiency compared to the original dQEq is achieved by iteratively minimizing a multidimensional function of the atomic partial charges using a gradient based method, which 
provides essentially the same results. This \textit{iterative} charge equilibration (iQEq) thus avoids the construction and solution of the extended Coulomb matrix, and therefore reduces the scaling from cubic to approximately quadratic. 

In this paper, we introduce the technical details and a benchmark application of the iQEq method, which we have implemented in the open-source software LAMMPS~\cite{P4473}, allowing for large-scale MD simulations of periodic systems using the 4G-HDNNP method. This new pair style can be used on multiple cores and is developed as an interface to the n2p2 library~\cite{P5603} for HDNNPs, which to date did not offer the option to run MD simulations based on 4G-HDNNPs. 
In the next section we provide a concise summary of the 4G-HDNNP method based on conventional dQEq. Then, we describe the iterative iQEq methodology and its implementation in LAMMPS. To illustrate the accuracy and performance of iQEq, we have carried out MD simulations of aqueous FeCl$_3$ solutions with different concentrations and compositions employing a recently reported 4G-HDNNP potential~\cite{P6302} for this system. Iron chloride solutions are challenging for MLPs, since the iron ions can adopt two different oxidation states in solution, Fe$^{2+}$ and Fe$^{3+}$. If data for both ions is combined in a single data set, in local second-generation and third-generation MLPs the oxidation state is ill-defined, since chloride ions may be located outside the local environments resulting in insufficient information to define the iron oxidation state. As shown in Ref.~\citenum{P6302}, 4G-HDNNPs are able to overcome this problem and to describe both, aqueous FeCl$_2$ as well as FeCl$_3$ solutions, correctly.

\section{Methods}\label{sec:method}
\subsection{4G-HDNNP with Direct Charge Equilibration}

\begin{figure}[!]
    \centering
    \includegraphics[width=0.45\textwidth]{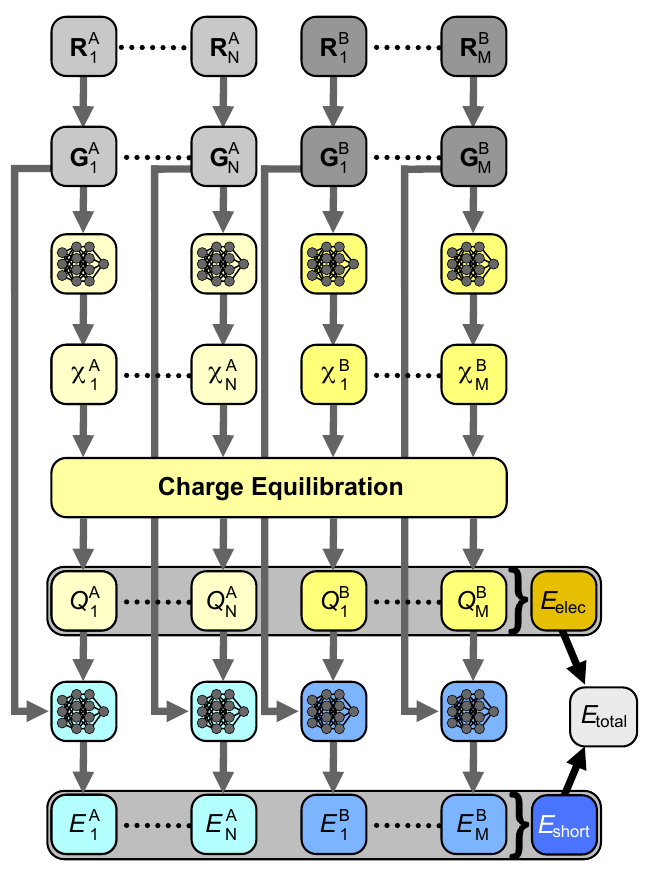}
    \caption{Schematic structure of the direct 4G-HDNNP method for a binary system~\cite{P5932}. The $\mathbf{R}_i$ are the Cartesian coordinate vectors of the $N$ atoms of element A and $M$ atoms of element B, which are first transformed to atom-centered symmetry function vectors $\mathbf{G}_i$ describing the atomic environments. These vectors then serve as inputs for the atomic neural networks predicting the atomic electronegativities $\chi_i$ that are used in a global charge equilibration step to yield the atomic charges $Q_i$. These charges determine the electrostatic energy $E_{\mathrm{elec}}$. Moreover, the $Q_i$ and $\mathbf{G}_i$ of each atom are used as input for a second atomic neural network providing its atomic energy $E_i$. The sum of all atomic energies $E_{\mathrm{short}}$ is then combined with the electrostatic energy to the total potential energy $E_{\mathrm{total}}$ of the system.}
    \label{fig:4g_diagram}
\end{figure}

A 4G-HDNNP~\cite{P5932} is a nonlocal MLP of the fourth-generation~\cite{P6018} containing atomic energies to represent local interactions and long-range electrostatic interactions based on structure-dependent atomic charges. In contrast to strictly local second-generation 2G-HDNNPs~\cite{P1174}, in which the atomic interactions are environment-dependent and smoothly truncated at a fixed cutoff radius~\cite{P2882}, the atomic energy contributions in 4G-HDNNPs also depend on the respective atomic charges that serve as global descriptors and are determined in a direct charge equilibration step~\cite{P1448} distributing the atomic charge density globally within the system. Therefore, the method is able to describe long-range charge transfer, which can give rise to local charge density changes throughout the system in response to very distant structural features or chemical reactions. The purpose of using the atomic charges as additional descriptors for the atomic energies is to take the effect of these charge density changes on local bonding in a consistent way into account.

The workflow of a 4G-HDNNP is shown schematically in Figure \ref{fig:4g_diagram}. Starting from the Cartesian coordinate vectors $\{\mathbf{R}_i\}$, first a vector $\mathbf{G}_i$ of atom-centered symmetry functions (ACSF)~\cite{P2882} is computed for each atom $i$, which is invariant with respect to translation and rotation of the system as well as permutation of like atoms and provides a local structure fingerprint of the atomic environment up to a cutoff radius $R_{\mathrm{c}}$. Then, a first set of atomic feed-forward neural networks (NN) is used to compute environment-dependent atomic electronegativities $\chi_i$, which allow to compute the atomic charges $Q_i$ in a global charge equilibration step. These charges are used for two purposes, first to calculate the electrostatic energy $E_{\mathrm{elec}}$ and second, together with the respective $\mathbf{G}$-vector as additional input neuron for a second atomic NN providing the atomic energy contributions $E_i$, which are added to yield $E_{\mathrm{short}}$. The total potential energy $E_{\mathrm{total}}$ of the system is then given by
\begin{eqnarray}
    E_{\mathrm{total}} = E_{\mathrm{elec}} + E_{\mathrm{short}} \quad .
\end{eqnarray}
The mapping from atomic electronegativities $\chi_i$ to atomic partial charges $Q_i$ via QEq is the key step of the 4G-HDNNP method. In the original 4G-HDNNP method~\cite{P5932} this is achieved by solving a system of linear equations, and here we will refer to this approach as direct dQEq, which is commonly used in different charge equilibration methods~\cite{P4419,P5880,P3350,P6829}.

The linear equations of dQEq are obtained as first derivatives of the charge equilibration energy 
\begin{equation} 
    E_\text{QEq} = E_\text{Coulomb} + \sum_{i=1}^{N_\text{atoms}} \chi_i Q_i + \frac{1}{2} J_i Q_i^2 \quad,
\end{equation}\label{e_qeq}
i.e., by minimizing this energy expression with respect to the atomic partial charges. Specifically, $E_\text{QEq}$ consists of the classical Coulomb energy $E_{\mathrm{Coulomb}}$ of the Gaussian-broadened atomic charges and a second-order Taylor expansion of the atomic energies in terms of the $Q_i$. $J_i$, which is an element-specific hardness to be optimized,
as well as the parameters of the NNs determining the electronegativities $\chi_i$, are obtained in a training process, in which they are adapted to reproduce reference charges from electronic structure calculations.
Often, Hirshfeld charges~\cite{P0416} are used for this purpose. However, the quality of the final potential does not significantly depend on the particular choice of partitioning method as the main purpose of the charges is the use as additional input features of the atomic neural networks. Moreover, long-range electrostatic interactions are efficiently screened in aqueous electrolytes while at short interatomic distances, i.e., within the cutoff radius of the ACSFs, any error in the electrostatic energy can be corrected by the atomic energies.

The energy $E_{\mathrm{QEq}}$ can be rewritten in matrix notation as 
\begin{equation} 
    E_\text{QEq} = \frac{1}{2} \mathbf{Q}^T \mathbf{A} \mathbf{Q} + \mathbf{Q}^T \bm{\chi},
\end{equation}\label{e_qeq2}
where the matrix $\mathbf{A}$ with dimensions $N_\text{atoms} \times N_\text{atoms}$ represents $E_{\mathrm{Coulomb}}$ as well as the atomic hardness. The elements of $\mathbf{A}$ have different forms for non-periodic and periodic systems:
\begin{itemize}
    \item Non-periodic system:
    \begin{equation}
        A_{ij} = 
        \begin{cases}
            J_i + \frac{1}{\sqrt{\pi} \sigma_i}, & \text{for } i = j \\
            \frac{\text{erf}\left(\frac{R_{ij}}{\sqrt{2} \gamma_{ij}}\right)}{R_{ij}}, & \text{for } i \neq j
        \end{cases}
    \end{equation}
    \item Periodic system (Ewald summation \cite{P0935}):
    \begin{equation}\label{eq:Amatrix}
    \begin{aligned}
        A_{ij} = \underbrace{\frac{4 \pi}{V} \sum_{\mathbf{k} \neq \mathbf{0}} \frac{e^\frac{-\eta^2|\mathbf{k}|^2}{2}}{|\mathbf{k}|^2} \cos \left[\mathbf{k}(\mathbf{R}_i - \mathbf{R}_j)\right]}_{=: A_{ij}^{\text{recip}}} \\ + 
         \begin{cases}
            J_i - \frac{2}{\sqrt{2 \pi} \eta} + \frac{1}{\sqrt{\pi} \sigma_i}, & \text{for } i = j \\
            \frac{\text{erfc}\left(\frac{R_{ij}}{\sqrt{2} \eta}\right)}{R_{ij}} - \frac{\text{erfc}\left(\frac{R_{ij}}{\sqrt{2} \gamma_{ij}}\right)}{R_{ij}}, & \text{for } i \neq j, R_{ij} < R_\text{cut}^\text{real}
        \end{cases}
    \end{aligned}
    \end{equation} 
\end{itemize}
The parameters $\sigma_i$ determine the widths of the Gaussians representing the atomic charges for the computation of $E_{\mathrm{Coulomb}}$ and $\gamma_{ij}=\sqrt{\sigma_i^2+\sigma_j^2}$, whereas $\eta$ is a hyperparameter defining the spatial extension of the real space part in the Ewald sum.
To obtain the atomic charges, $E_{\mathrm{QEq}}$ is minimized by solving 
\begin{equation} \label{E_qeq_solution}
    \frac{\partial{ E_{\mathrm{QEq}}}}  {\partial Q_i} = 0
\end{equation}
with the constraint $\sum_{i} Q_i = Q_{\mathrm{tot}}$ for each atom $i$. This total charge conservation constraint is enforced by a Lagrange multiplier $\lambda$ and leads to an extended matrix $\mathbf{A'}$ with dimension $(N_\text{atoms}+1) \times (N_\text{atoms}+1)$,
\begin{eqnarray}\label{eq:matrix}
	\left( \begin{array}{ccc|c}
		  &         &   & 1      \\
		  & A_{i,j} &   & \vdots \\
		  &         &   & 1      \\ \hline
		1 &  ...    & 1 & 0      \\
	\end{array}\right)
	\left(\begin{array}{c}
		Q_{1}   \\
		\vdots  \\
		Q_{N_{\mathrm{atoms}}}   \\ \hline
		\lambda \\
	\end{array}\right)
	=
	\left(\begin{array}{c}
		-\chi_{1} \\
		\vdots    \\
		-\chi_{N_{\mathrm{atoms}}} \\ \hline
		Q_{\mathrm{tot}}   \\
	\end{array}\right) 
\end{eqnarray}
or, introducing a solution vector $\bm{b}$ representing the right hand side of Eq.~\ref{eq:matrix},
\begin{equation} \label{eq:eqeq_solution}
    \mathbf{A'} \mathbf{Q'} = \bm{b} \quad .
\end{equation}
This equation can then be solved to determine the atomic partial charges by calculating the inverse of the matrix $\mathbf{A'}$. 
Once the atomic charges have been obtained, atomic forces can be calculated as described in Ref.~\citenum{P5932}. 

\subsection{Iterative Charge Equilibration}\label{sec:modified_method}

To improve the computational performance of 4G-HDNNPs, an increased efficiency of the charge equilibration step is central. We address this problem by replacing the cubically scaling dQEq method with an \textit{iterative} iQEq algorithm, which performs a multidimensional function minimization for calculating the atomic charges and force components. Solving a system of linear equations iteratively via gradient-based optimizations such as the conjugate gradient algorithm has been proposed in the literature before~\cite{P6683} and it has been shown that the scaling of the solution can be enhanced significantly. Recently, also Gubler et al. \cite{P6789} discussed this issue and showed that an iterative solution using a rapidly converging conjugate gradient method in combination with a particle mesh solver leads to substantially improved scaling with respect to the number of atoms in the system.

In dQEq, constructing the matrix $\mathbf{A'}$ scales quadratically with the number of atoms, while the solution of Eq.~\ref{eq:matrix} exhibits cubic scaling.
In the iQEq method, this solution is replaced by a multidimensional function minimization scheme exploiting that the charge equilibration energy $E_\mathrm{QEq}$ is a multidimensional function of the atomic charges 
\begin{equation}\label{eq:E_QEq}
        E_\text{QEq} = E_\text{QEq} (Q_1,\ldots,Q_{N_\text{atoms}}) \quad. 
\end{equation} 
Minimizing this function using a gradient-based approach in combination with a total charge constraint
\begin{equation}
        Q_\text{tot} = \sum_{i=1}^{N_\text{atoms}} Q_i
\end{equation}
yields the atomic charge vector. The gradient vector of the charge equilibration energy with respect to the charge vector $\frac{\partial E_\text{QEq}}{\partial Q_i}$ can be obtained for non-periodic and periodic systems as
\begin{itemize}

     \item Non-periodic system:
    
    \begin{equation}
         \frac{\partial{ E_{\mathrm{QEq}}}}{\partial Q_i} = \frac{Q_i}{\sigma_i \sqrt{\pi}} + \chi_i + J_i Q_i + \sum_{j \neq i }^{N_\text{atoms}} \frac{\text{erf}\left(\frac{r_{ij}}{\sqrt{2} \gamma_{ij}}\right)}{r_{ij}}Q_j
    \end{equation}

    \item Periodic system:
    
    \begin{equation}\label{eq:EQEq_gradient}
    \begin{aligned}
         \frac{\partial{ E_{\mathrm{QEq}}}}{\partial Q_i} = \chi_i + J_i Q_i +    \frac{4 \pi}{V}  \sum_{\vec{\mathbf{k}} \neq \mathbf{0}} \frac{e^\frac{-\eta^2|\vec{\mathbf{k}}|^2}{2}}{|\vec{\mathbf{k}}|^2} \mathrm{Re}\left(S(\mathbf{k})e^{i\mathbf{k}\mathbf{r}_i}\right)  + \\
         \sum_{j }^{N_\text{neig}} \frac{\text{erfc}\left(\frac{R_{ij}}{\sqrt{2} \eta }\right) - \text{erfc}\left(\frac{R_{ij}}{\sqrt{2} \gamma }\right)}{R_{ij}} Q_j + \frac{Q_i}{\sqrt{\pi} \sigma _i} - \frac{2 Q_i}{\sqrt{2\pi} \eta} 
    \end{aligned}
    \end{equation} 
\end{itemize} 
where $S(\mathbf{k})$ is the structure factor defined as 
\begin{equation}\label{eq:structure_factor}
       S(\mathbf{k}) =   \sum_{i=1}^{N_\text{atoms}} Q_i \text{exp}(i\mathbf{k} \cdot \mathbf{r}_i).
\end{equation} 
Using the gradient vector, the QEq energy can then be minimized using any gradient-based algorithm. This minimization is an iterative process and the number of iterations depends on the selected gradient algorithm and tolerance values. Two different tolerance parameters need to be selected in the iQEq approach: one is the tolerance for the line minimization and the other is the tolerance for the convergence (i.e., gradient tolerance). The former sets the accuracy for the line search process within the minimization algorithm. During each iteration, the algorithm searches along a particular direction defined by the gradient or other optimization criteria to find a step size that minimizes the function in that direction. The latter sets the convergence criterion for the entire minimization process. It ensures that the norm of the gradient falls below a certain threshold, meaning that a local minimum has been reached. In this study, these tolerances are set to 10$^{-2}$~eV and 10$^{-5}$~eV/e, respectively, and in general the values to be chosen depend on the desired accuracy and performance as discussed below. It has been observed that the iterative  solution significantly improves the scaling of the method with system size, making it attractive for simulations of large systems. The detailed performance comparison between iQEq and dQEq methods is provided in Section \ref{sec:results}. 

After calculating the atomic charges, the next step is to compute the electrostatic energy and forces. For the computation of the forces in principle the partial derivatives of the charges with respect to all atomic coordinates are required. However, this approach is computationally demanding as it requires solving a system of linear equations for each force component, i.e., $3 \times N_{\mathrm{atoms}}$ systems of linear equations, since the matrix elements are different for each component. To avoid this demanding step, an efficient Lagrange formalism is utilized that allows for the calculation of forces by solving only a single system of linear equations as derived in the supporting information of Ref.~\citenum{P5932}.  

First, to simplify the force calculation, we introduce an auxiliary function $L$ defined as
\begin{equation}\label{eq:force_trick1}
L = E_{\text{total}} + \sum_{i=1}^{N_{\text{atoms}}+1} \lambda_i \left( \sum_{j=1}^{N_{\text{atoms}}+1} A'_{ij} Q'_j - b_i \right) \quad , 
\end{equation} 
where $E_{\text{total}}$, i.e., $E_{\text{elec}}$ + $E_{\text{short}}$, depends on both the atomic coordinates and the charges, which themselves are functions of the atomic coordinates. Here $\sum_{j=1}^{N_{\text{atoms}}+1} A'_{ij} Q'_j - b_i$ represent the differences between the left-hand side and the right-hand side of Eq. \ref{eq:eqeq_solution} that is used to compute the charges $Q_i$, ensuring that $L$ is always equal to $E_{\text{total}}$. The multipliers $\lambda$ can be chosen such that the partial derivatives $\frac{\partial L}{\partial Q_{i}}$ are zero,
\begin{equation}\label{eq:force_trick2}
\frac{\partial L}{\partial Q'_i} = \frac{\partial E_{\text{total}}}{\partial Q'_i} + \sum_{j=1}^{N_{\text{atoms}}+1} A'_{ij} \lambda_j = 0 \quad . 
\end{equation}
This results in a system of linear equations that can be solved as 
\begin{equation}\label{eq:force_trick3} 
\sum_{j=1}^{N_{\text{atoms}}+1} A'_{ij} \lambda_j = -\frac{\partial E_{\text{total}}}{\partial Q'_i} \quad . 
\end{equation} 
Since $\mathbf{A}'$ is a symmetric matrix, the computational complexity is reduced significantly.

The derivative $\frac{dL}{dR_{\alpha}}$ corresponds to $\frac{dE_{\mathrm{total}}}{dR_{\alpha}}$, which is the derivative of the total energy with respect to the atomic coordinates, i.e., an atomic force. This derivative can then be written as 
\begin{align}\label{eq:force_trick4}
&\frac{dE_{\text{total}}}{dR_\alpha} = \frac{\partial E_{\text{total}}}{\partial R_\alpha} + 
\sum_{i=1}^{N_{\text{atoms}}+1} \frac{\partial E_{\text{total}}}{\partial Q'_i} 
\frac{\partial Q'_i}{\partial R_\alpha} \nonumber \\
&+ \sum_{i=1}^{N_{\text{atoms}}+1} \lambda_i \left( \sum_{j=1}^{N_{\text{atoms}}+1} 
\frac{\partial A'_{ij}}{\partial R_\alpha} Q'_j + \sum_{j=1}^{N_{\text{atoms}}+1} 
A'_{ij} \frac{\partial Q'_j}{\partial R_\alpha} - \frac{\partial b_i}{\partial R_\alpha} \right) 
\end{align}
By rearranging this equation and eliminating terms that are zero by definition of $\lambda$, the final expression for the atomic forces is obtained as
\begin{equation}
\frac{dE_{\text{total}}}{dR_\alpha} = \frac{\partial E_{\text{total}}}{\partial R_\alpha} + \sum_{i=1}^{N_{\text{atoms}}+1} \lambda_i \left( \sum_{j=1}^{N_{\text{atoms}}+1} \frac{\partial A'_{ij}}{\partial R_\alpha} Q'_j - \frac{\partial b_i}{\partial R_\alpha} \right) \label{eq:force_trick5}
\end{equation}
This approach significantly reduces the computational cost, making it an efficient method for calculating atomic forces in large-scale simulations. In the original 4G-HDNNP method, a second matrix solution was necessary to obtain $\lambda_i$ and to calculate the electrostatic forces. The matrix solution for $\lambda_i$ in Eq.\ \ref{eq:force_trick3} is replaced by an iterative solution using a gradient-based function minimization. Similar to Eq.\ \ref{eq:E_QEq}, the fact that Eq.\ \ref{eq:force_trick5} can be reformulated as a multidimensional function of $\lambda_i$ can be exploited to yield  
\begin{equation}
\frac{dE_{\text{total}}}{dR_\alpha} = \frac{\partial E_{\text{total}}}{\partial R_\alpha} + f(\lambda_i).
\end{equation}\label{eq:dE_QEq_lambda}

\subsection{Implementation}

\begin{figure}[!]
    \centering
    \includegraphics[width=0.5\textwidth]{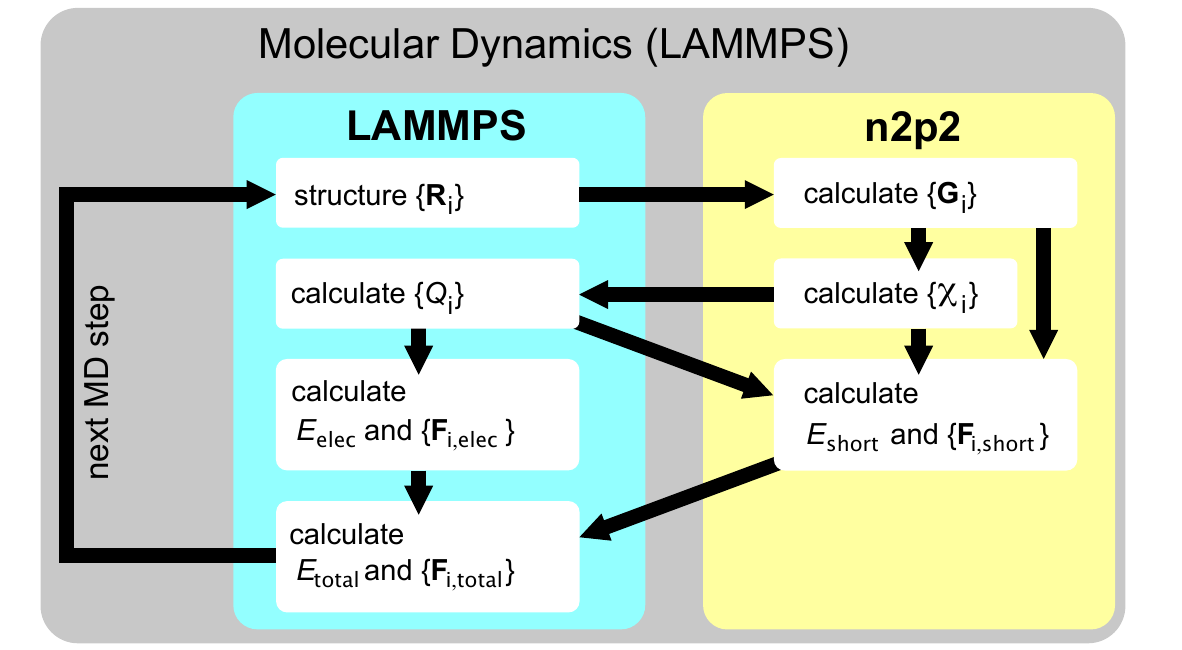}
    \caption{Schematic workflow of the 4G-HDNNP implementation within the framework of LAMMPS and n2p2. The global iQEq is used in the calculation of the charges $\{Q_i\}$, and of the electrostatic energy and forces. The arrows illustrate the flow of data and computational steps, where ${\mathbf{R}_i}$ are Cartesian coordinates, $\{\mathbf{G}_i\}$ the atom-centered symmetry function vectors, $\{\chi_i\}$ the atomic electronegativities, $\{E_i\}$ the atomic energies and $\{\mathbf{F}_i\}$ the atomic force vectors.} 
    \label{fig:4G_interface}
\end{figure}

We have implemented the iQEq method in the LAMMPS software~\cite{P4473} building upon the n2p2 library~\cite{P5603}, which provides the HDNNP method in LAMMPS as a pair style. The primary objective of the current implementation was to extend the n2p2 library to include 4G-HDNNPs and to parallelize the global charge equilibration and force calculation steps to enable running large-scale MD simulations. An overview of the workflow of the 4G-HDNNP interface is given in Figure \ref{fig:4G_interface}. 

In LAMMPS, all computations other than atomic forces are carried out by entities known as \texttt{fix} or \texttt{compute}, executing specific tasks at designated points, i.e., before or after the force calculation, during an MD timestep. The iQEq process is implemented as a \textit{pre-force} \texttt{fix} style, labeled as \texttt{FixHDNNP},
in order to make atomic charges available in the second-stage, where the short-range forces and energies are predicted using atomic charges as additional input neurons. This \texttt{fix} interfaces to n2p2 like the \texttt{PairHDNNP} pair style and initiates the execution of the initial set of neural networks to predict the atomic electronegativities. Atomic charges are calculated in parallel in this \texttt{fix} style through the iQEq method as described in Section \ref{sec:modified_method}. For this calculation, the multidimensional function minimization algorithm in the open-source GNU Scientific Library (GSL) \cite{galassi2002gnu} has been used. Specifically we found the best performance for the BFGS-2 algorithm \cite{fletcher2000practical}. After calculating the atomic charges iteratively, the next step involves the calculation of the electrostatic contributions to the forces. For this, a new \texttt{KSpaceHDNNP} class was inherited from the native \texttt{KSpace} class in LAMMPS. This class is equipped with the necessary data structures and routines to calculate the electrostatic force contributions, thereby facilitating the incorporation of long-range interactions.
The final steps of the 4G-HDNNP method are realized in the \texttt{PairHDNNP} class, which inherits from the \texttt{Pair} class and acts as a pair style. This class interfaces to the n2p2 library and forwards the computed atomic charges to the n2p2 library. Like the first set of atomic NNs, the second set of atomic NNs is evaluated on the n2p2 side and predicts the short-range atomic energies and forces. These are then transferred back to LAMMPS and added to the electrostatic contributions that have already been calculated. Once the complete atomic force vectors have been determined, the equations of motion are integrated and the next MD time step of the MD simulation is executed.

For an efficient parallelization of the charge equilibration method, the MPI (Message Passing Interface) library \cite{mpi40} is used to manage communications across processors. The GSL multidimensional function minimization algorithm \cite{galassi2002gnu} is employed to run the optimization process, exploiting the fact that the charge equilibration energy $E_\mathrm{QEq}$ in Eq.\ \ref{eq:E_QEq} can be expressed as a multidimensional function of atomic charges $Q_i$. For the minimization of $E_\mathrm{QEq}$, the gradient vector $\frac{\partial E_\mathrm{QEq}}{\partial Q_i}$ in Eq.\ \ref{eq:EQEq_gradient} needs to be calculated as well. Since each processor contains only a fraction of the global atom vector in LAMMPS (i.e., ``local atoms''), one can distribute the calculation of $E_\mathrm{QEq}$ and $\frac{\partial E_\mathrm{QEq}}{\partial Q_i}$ using only the charges and position vectors of these local atoms. Two key functions are necessary within this framework: one for calculating the charge equilibration energy as a scalar and another for computing the gradient vector of the charge equilibration energy with respect to the charge vector. Both $E_\mathrm{QEq}$ and $\frac{\partial E_\mathrm{QEq}}{\partial Q_i}$ have one real space, one reciprocal space and one self-correction term as given in Eqns. \ref{eq:Amatrix} and \ref{eq:EQEq_gradient}. No additional communication is necessary for real space contributions and self-correction terms as all local atoms have neighbor lists attached to them. However, an additional communication step is necessary for the calculation of reciprocal space energy and gradient contributions on each processor. This is because of global structure factors in Eq.\ \ref{eq:structure_factor} that can be calculated by looping over all atoms within the system. After defining these functions, 
local contributions are communicated from each processor. Specifically, the local energy contributions and the local gradient vectors are communicated and aggregated to obtain the global energy and gradient values $E_\mathrm{QEq}$ and $\frac{\partial E_\mathrm{QEq}}{\partial Q_i}$. This communication ensures that each processor has the cumulative global quantities and then the gradient-based optimization of GSL finds the charge vector $Q_i$ with the selected gradient tolerance across all processors. The same parallelization approach is applied for the calculation of the parameters $\lambda_i$ necessary for the efficient electrostatic force calculation. Local energies and gradient vectors are computed and similarly communicated across processors  
before the gradient-based optimization step. 

\section{Results}\label{sec:results}

\subsection{Performance}

\begin{figure}[!]
    \centering
    \includegraphics[width=0.45\textwidth]{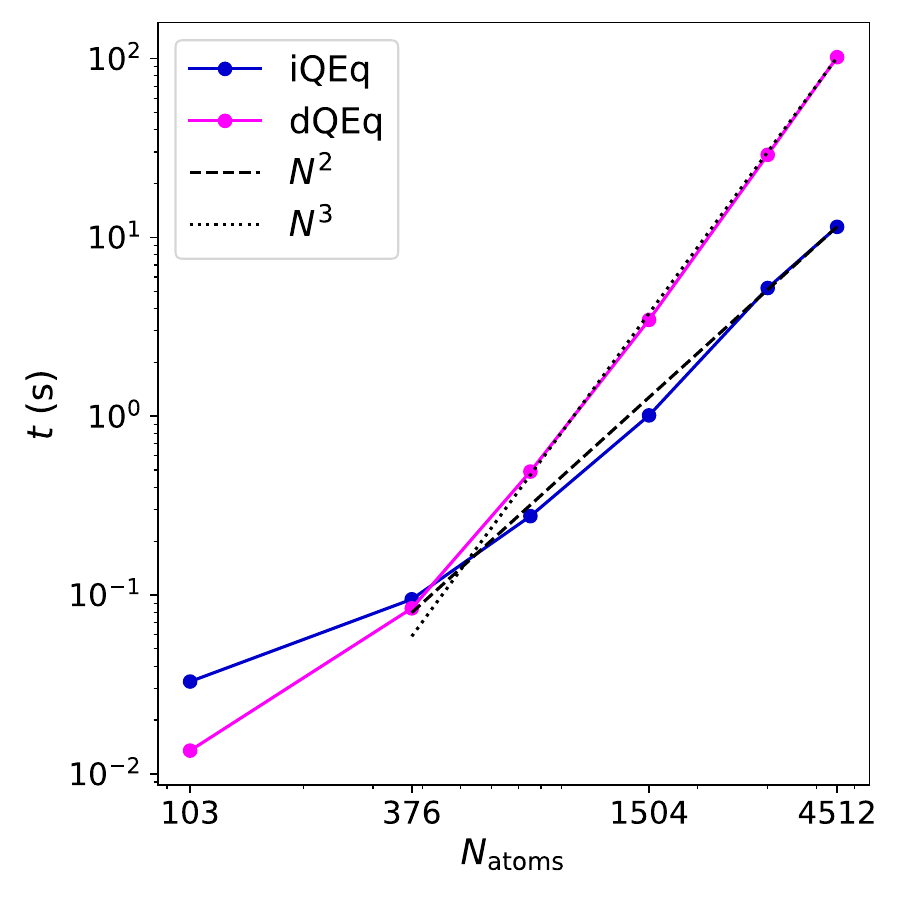}
    \caption{Calculation times for the charge determination in the iQEq and dQEq methods as a function of system size on a single CPU core (Intel i7-12700). Six different systems have been used consisting of 103, 376, 752, 1505, 3008 and 4512 atoms, respectively. Reference lines for $N^2$ and $N^3$ scaling are included to illustrate the performance of both approaches. 
} 
    \label{fig:dQEq_vs_iQEq}
\end{figure}

In a first step, we investigate the computational efficiency and scaling behavior of the iQEq implementation in LAMMPS with respect to the original dQEq matrix solution for an increasing number of atoms and assess how the method benefits from multiple CPU cores in parallel runs. 

The original 4G-HDNNP approach computes atomic charges by the dQEq method and thus involves building and inverting a $(N_\text{atoms}+1) \times (N_\text{atoms}+1)$ $\mathbf{A'}$ matrix that holds information about every atom in the system. As the number of atoms increases, this approach becomes computationally demanding, as the costs of constructing and inverting the matrix scale quadratically and  cubically, respectively, with the size of the system. 

The modified 4G-HDNNP method uses the iQEq scheme introduced in Section \ref{sec:modified_method} and is expected to show a higher computational performance. The timings of the dQEq and iQEq methods on a single CPU core 
are compared in Fig.~\ref{fig:dQEq_vs_iQEq} using six different cubic boxes of water each of which contains a single formula unit of FeCl$_3$ resulting in overall 103, 376, 752, 1504, 3008, and 4512 atoms, respectively. 
For small systems up to about 400 atoms the dQEq method exhibits smaller computation times, 
but as the number of atoms increases, the iQEq method benefits from a quadratic scaling behavior. The matrix-based dQEq method scales cubically for larger system sizes, resulting in a steep increase in computational cost. These results are consistent with the findings of Gubler et al. \cite{P6789} and underline the advantages of using iterative solvers for the charge equilibration step in 4G-HDNNPs.

When measuring the CPU times of the iQEq method it is important to note that a tolerance value has to be chosen defining the accuracy of the iterative charge determination. Moreover, a maximum number of iterations for terminating the iterative minimization needs to be specified. 
These parameters need to be tested for a given system and depending on the choice of these parameters the absolute timings of the iQEq algorithm are necessarily different.
For our particular analysis, we choose 10$^{-5}$ eV/e as the gradient tolerance and used up to 15 iterations. As discussed in detail below, the required settings depend on the desired accuracy, and here we achieve a typical convergence up to the fifth decimal for both atomic charges and forces in units of e and eV/\AA{}, respectively. 

The parallelization speedup of the iQEq implementation has been tested in the range from 1 to 32 CPU cores for cubic FeCl$_3$-water boxes using four different system sizes. The time required to complete the simulation is measured for each core count, and the speedup factor is then calculated as the ratio of the computation time on a single core to the computation time with a given number of cores. A linear speedup would correspond to a perfect scaling without any notable overhead due to communication between processes, with the computation time being inversely proportional to the number of cores. In real calculations the speedup curve starts to saturate after a certain number of cores.
In LAMMPS, this saturation behavior in general strongly depends on the number of atoms per core. There is an optimal number of atoms per core when running MD simulations and this optimal number depends on several factors like hardware and employed potential. 
Thus, it is usually required to test different numbers of cores to determine the optimal setup for a specific case. 

\begin{figure}[!]
    \centering
    \includegraphics[width=0.45\textwidth]{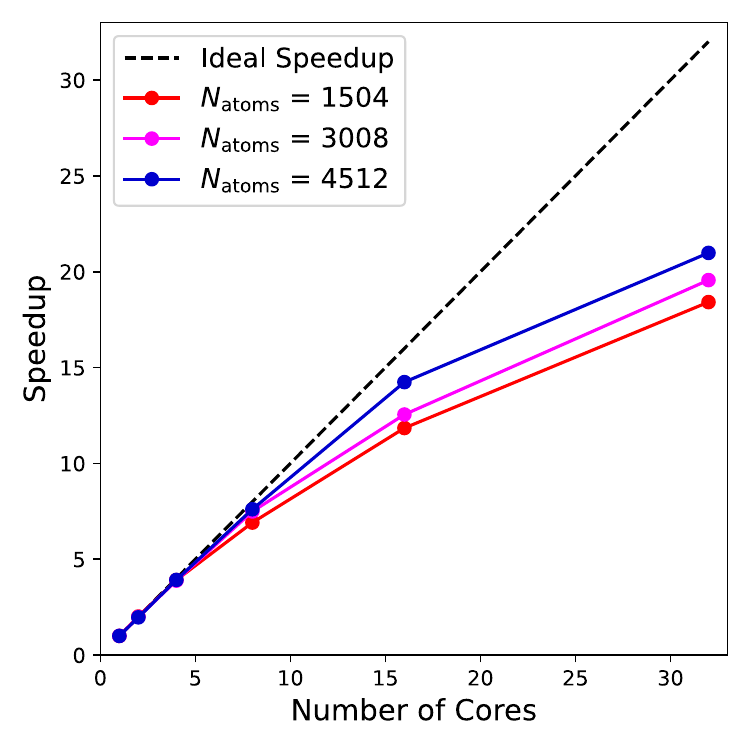}
    \caption{Speedup curves for three different system sizes with $N_{\mathrm{atoms}}$ = 1504, 3008 and 4512. The iQEq execution times are measured on 1, 2, 4, 8, 16, and 32 CPU cores (Intel Xeon 4514Y CPU, Dual-CPU system). The speedup is calculated by averaging the iQEq execution times of multiple runs for each core count needed to determine the charges, relative to the baseline performance with a single core. The ideal speedup line is included for reference, representing a linear increase in performance as the number of cores is increased. 
    }
    \label{fig:speedup}
\end{figure}

Figure \ref{fig:speedup} shows the speedup of the 4G-HDNNP iQEq step for various system sizes with respect to the ideal speedup line. For a given number of cores the parallelization efficiency improves significantly with increasing number of atoms in the system. For the smallest system, the speedup curve deviates most rapidly from the ideal line as a consequence of the small number of atoms per core and the resulting relatively high impact of communication between processes. Therefore, it is essential to balance the number of cores with the size of the system to achieve the best parallel performance for a given setup. The fact that the speedup curve gradually saturates for all system sizes is attributed to the underlying QEq approach, which requires inter-core communications during the iterative charge optimization.

\subsection{Accuracy}

\begin{figure}[!]
    \centering
    \includegraphics[width=0.48\textwidth]{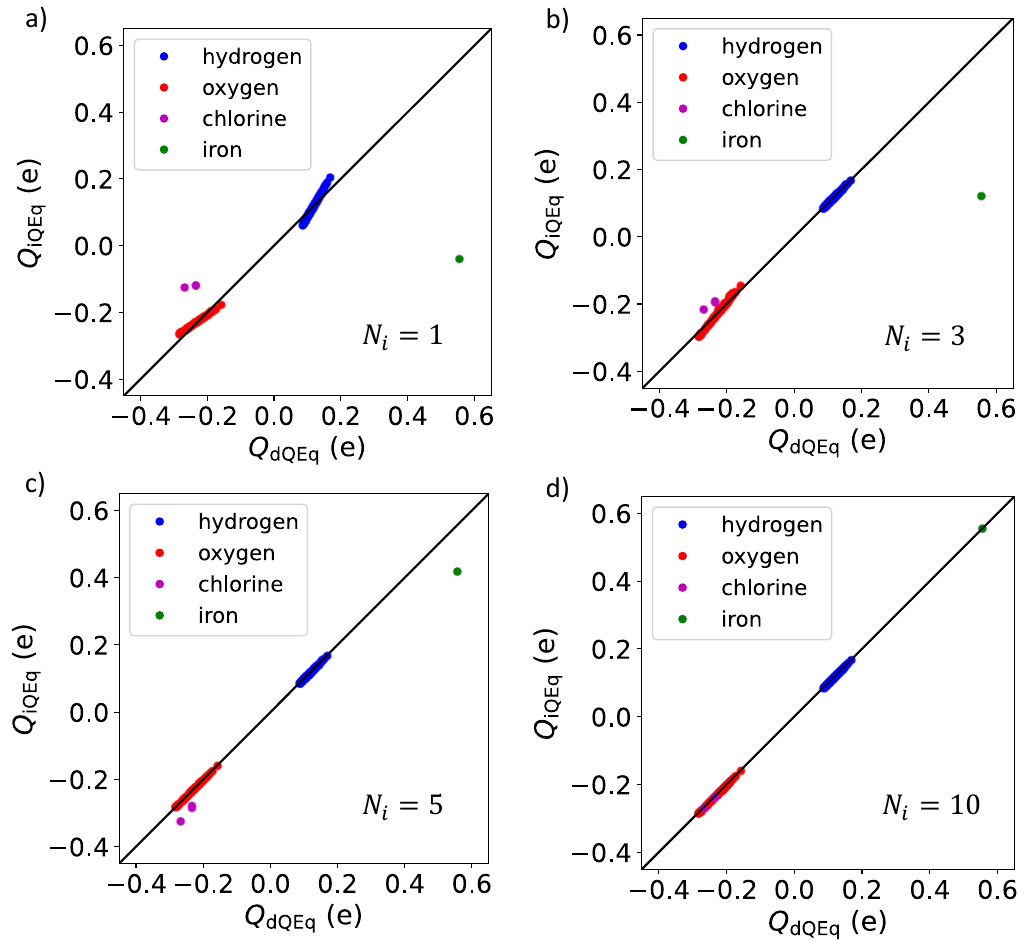}
    \caption{Comparison of the charges $Q_{\mathrm{iQEq}}$ predicted by the iterative iQEq method after $N_i=$ 1, 3, 5 and 10 iterations (panels a, b, c, and d, respectively) with the charges $Q_{\mathrm{dQEq}}$ predicted by the direct dQEq method for a system of FeCl$_3$ in water containing overall 376 atoms.} 
    \label{fig:convergence_charge}
\end{figure}

\begin{figure}[!]
    \centering
    \includegraphics[width=0.48\textwidth]{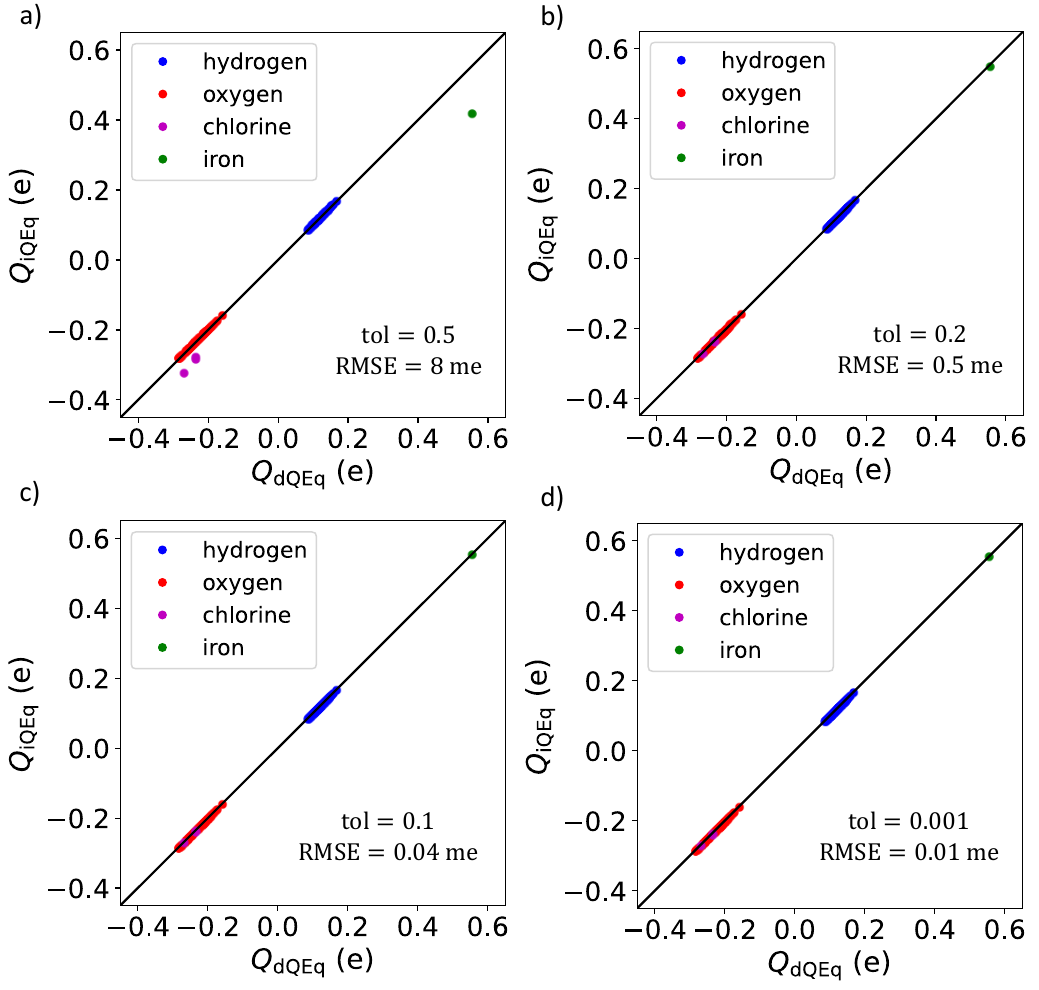}
    \caption{Comparison of the charges $Q_{\mathrm{iQEq}}$ predicted by the iterative iQEq method with gradient tolerance being equal to 0.5, 0.2, 0.1 and 0.001 eV/e (panels a, b, c, and d, respectively) with the charges $Q_{\mathrm{dQEq}}$ predicted by the direct dQEq method for a system of FeCl$_3$ in water containing overall 376 atoms. The root mean squared errors of the $Q_{\mathrm{iQEq}}$ with respect to the $Q_{\mathrm{dQEq}}$ are shown in the inset.} 
    \label{fig:convergence_charge_tol}
\end{figure}

\begin{figure}[!]
    \centering
    \includegraphics[width=0.48\textwidth]{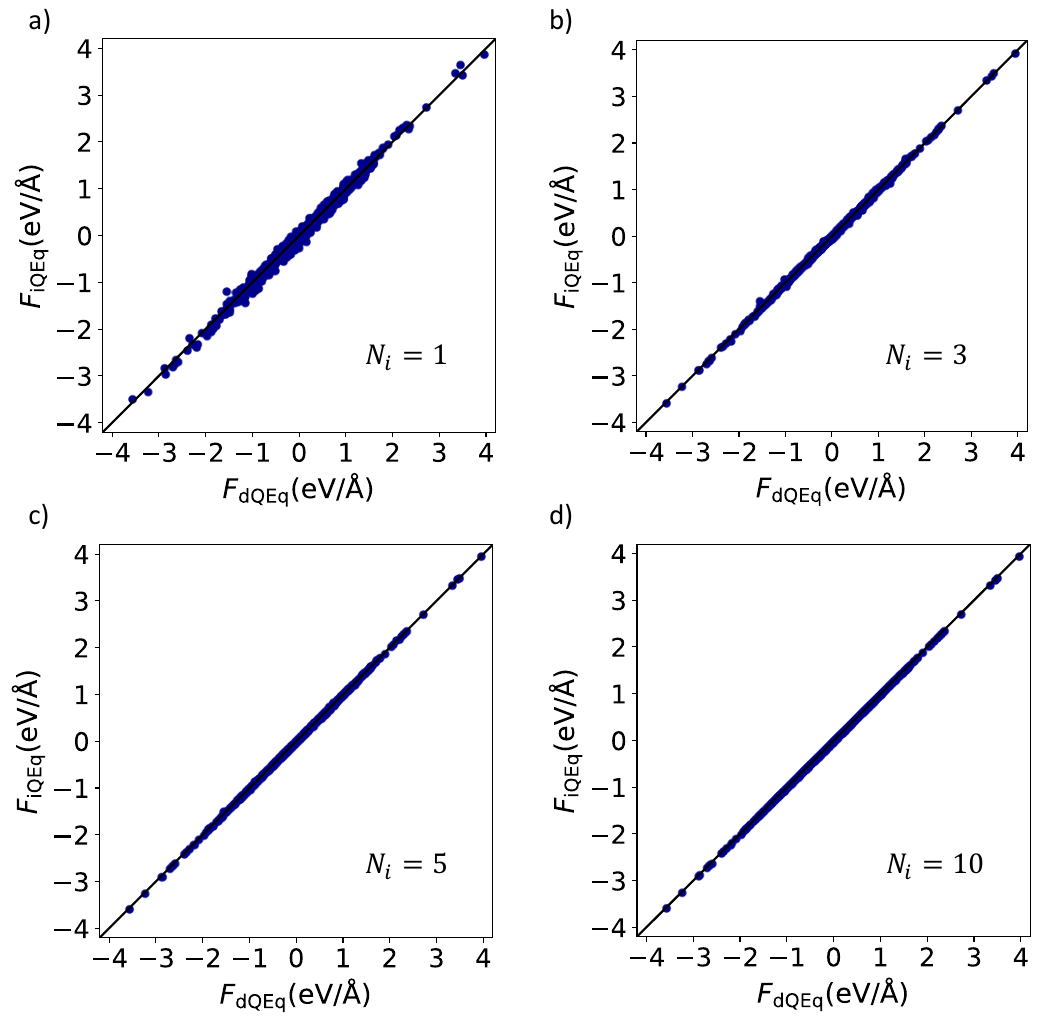}
    \caption{Comparison of total atomic force components $F_{\mathrm{iQEq}}$ predicted by the iterative iQEq method after $N_i=$ 1, 3, 5 and 10 iterations (panels a, b, c, and d, respectively) with the atomic force components $F_{\mathrm{dQEq}}$ predicted by the direct dQEq method for a system of FeCl$_3$ in water containing overall 376 atoms.} 
    \label{fig:convergence_force}
\end{figure}

In this section, the accuracy of the iQEq 4G-HDNNP method is assessed by comparing the calculated atomic charges and forces to those obtained from the original dQEq matrix-based solution. Both, the charges and forces, are derived iteratively in the new method, where the charges are computed through a multidimensional function minimization scheme and the forces require a second iteration cycle in which the parameter $\lambda$ is calculated as described in Section \ref{sec:modified_method}. It is important to make sure that both quantities converge to the desired accuracy for the predefined number of iterations $N_i$, which is a parameter that needs to be specified by the user. 

The effect of the number of iterations $N_i$ on the convergence of the atomic charges is illustrated in Figure \ref{fig:convergence_charge}, where charge predictions made by the iterative iQEq method are compared to those from the original direct dQEq method for a cubic FeCl$_3$-water box containing 376 atoms. To assess the convergence behavior, a series of correlation plots is provided, showing results for 1, 3, 5, and 10 iterations. As observed, the predicted charges increasingly align with the dQEq values as the number of iterations increases. Our findings indicate that within 10 iterations the iterative solution provides atomic charges that match the matrix solution with an RMSE of 0.0132~me  for the FeCl$_3$-water system. 

A second parameter that needs to be defined is the gradient tolerance as described in Section \ref{sec:modified_method}. The effect of this parameter on the convergence of the atomic charges is  shown in Figure \ref{fig:convergence_charge_tol}, where again charge predictions made by the iQEq method are compared to the dQEq values for the same system. Four different tolerances have been tested, which are 0.5, 0.2, 0.1 and 0.001 eV/e. We find that the iQEq predictions already converge to the dQEq results for tol = 0.1~eV/e. Moreover, we observed that a tighter convergence threshold (such as tol = 10$^{-5}$ eV/e) might be required for achieving an RMSE of 0.01~me. Therefore, we recommend using a gradient tolerance of 10$^{-5}$ eV/e to avoid the accumulation of errors that can lead to extrapolations and unstable trajectories. 

Similarly, the convergence of the total atomic forces, i.e., the sum of the short-range and electrostatic parts, with respect to $N_i$ is shown in Figure \ref{fig:convergence_force}, where force predictions from the 4G-HDNNP method that utilizes the iterative iQEq approach are compared to the original method that utilizes the direct dQEq approach for the same FeCl$_3$-water system. A series of correlation plots is presented for 1, 3, 5, and 10 iterations illustrating the convergence behavior of the forces. The predicted forces converge to the reference values as the number of iterations increases. Already after 5 iterations the forces predicted by the iterative approach match within 0.001 eV/\AA{} those from the direct solution, confirming the accuracy of the iterative solution. The RMSE of forces after 10 iterations is 0.059 meV/\AA{}.

\subsection{Molecular Dynamics Simulations}

To further validate the iQEq 4G-HDNNP method, a 500 ps long MD simulation has been performed in the $NVE$ ensemble with a FeCl$_3$-water system containing 376 atoms using the LAMMPS 4G-HDNNP interface. The gradient tolerance parameter  of the iterative solver has been selected as 10$^{-5}$ eV/e, whereas the maximum number of iterations has been set to 30 although we find that the method usually converges much earlier. One goal of this simulation has been to test the long-term stability of the implemented algorithm. Moreover, the FeCl$_3$ system has been selected to assess the ability of the method to accurately predict atomic charges given that the 4G-HDNNP has been trained to a dataset containing iron atoms in the Fe$^{2+}$ as well as Fe$^{3+}$ oxidation states~\cite{P6302}. The aim of including structures of both oxidation states during training has been to challenge the 4G-HDNNP method with information that cannot be captured by the short-range part of the method alone, since the iron oxidation state is defined by the number of chloride counter ions, which might be located outside the local environments of the Fe atoms. Instead, a global QEq approach is needed for this purpose~\cite{P6302}. 

\begin{figure}[!]
    \centering
    \includegraphics[width=0.45\textwidth]{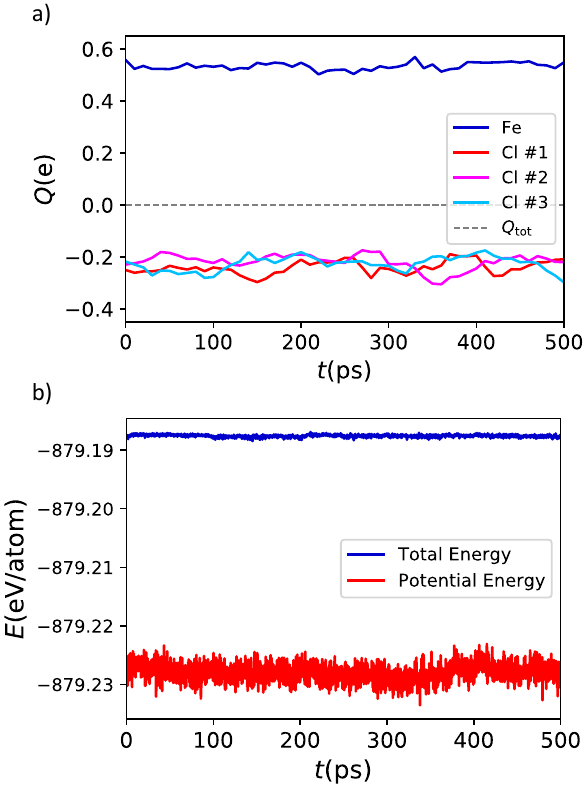}
    \caption{Time evolution of a) atomic charges for the Fe and Cl ions and b) the total and potential energy for a 500 ps long FeCl$_3$-water system trajectory containing 376 atoms. The simulation has been performed in the $NVE$ ensemble using the iQEq 4G-HDNNP method after a 5 ps long $NVT$ equilibration at 300 K. The 4G-HDNNP has been trained on a mixed dataset containing both Fe$^{2+}$ and Fe$^{3+}$ oxidation states~\cite{P6302}. 
    } 
    \label{fig:method_traj}
\end{figure}

The time evolution of the atomic charges for the Fe and Cl ions along the trajectory is presented in Fig. \ref{fig:method_traj}a. The predicted charges remain consistent with the expected values for all ions throughout the simulation. We note that the numerical values of these charges correspond to the values of DFT Hirshfeld charges \cite{P0416}, which have been used in constructing the 4G-HDNNP.  
To further assess the stability of the simulation, the time evolution of the potential energy and of the total, i.e., kinetic and potential, energy is examined. As shown in Fig. \ref{fig:method_traj}b, both the total and potential energies fluctuate slightly around stable mean values indicating that the system has reached an equilibrium and remains stable throughout the simulation. This demonstrates that the iQEq method does not only accurately predict charges but also conserves energy over time, which is crucial for long-term stability in MD simulations. The ability of the iQEq method to model complex ionic systems over long MD trajectories, even when trained on a mixed dataset containing multiple oxidation states, demonstrates its robustness of the 4G-HDNNP method for use within LAMMPS.

\section{Conclusion}

In this paper, a modified version of the fourth-generation high-dimensional neural network potential (4G-HDNNP) has been presented that incorporates an iterative charge equilibration (iQEq) method representing an efficient alternative to the direct matrix-based solution (dQEq). The modified charge equilibration algorithm has been implemented in the open source molecular dynamics software LAMMPS as a new pair style using the n2p2 library as an interface.

We showed that the iQEq method provides atomic charges and forces converging to the respective values of the dQEq method. While conserving the accuracy of the original method, the iQEq approach offers significant computational benefits, particularly for large-scale systems by avoiding the need to solve a global matrix with cubic scaling at every step. This improvement in computational efficiency is crucial for enabling the use of the 4G-HDNNP in large-scale molecular dynamics simulations while maintaining the accuracy necessary for complex chemical environments, and might also be of benefit for other similar fourth-generation machine learning potentials~\cite{P6829}. The implementation of the iQEq method into LAMMPS is a significant step forward in the application of the 4G-HDNNP method for large-scale, long-range interaction modeling in molecular dynamics simulations. In future work, further performance gains could be expected by replacing the Ewald sum with a particle mesh solver~\cite{P6789}.

\begin{acknowledgments}
We are grateful for support by the Deutsche Forschungsgemeinschaft (DFG) (BE3264/16-1, project number 495842446 in priority program SPP 2363 ``Utilization and Development of Machine Learning for Molecular Applications – Molecular Machine Learning'') and under Germany's Excellence Strategy – EXC 2033 RESOLV (project-ID 390677874). This research was funded in part by the Austrian Science Fund (FWF) 10.55776/F81. 
Discussions with Gunnar Schmitz are gratefully acknowledged.
\end{acknowledgments}

\section*{Data Availability Statement}
The data that support the findings of this study are available from the
corresponding author upon reasonable request.
The implementation of the iQEq method is available as open-source software at https:\///\/compphysvienna.github.io\//n2p2\/.

\bibliography{literature}

\end{document}